\documentstyle[twocolumn,aps,epsfig]{revtex}
\begin{document}
\draft
\title{Effect of localized holes on the  long-range order in
bilayer antiferromagnets}
\author{I.~Ya.~Korenblit, Amnon Aharony, and O.~Entin-Wohlman}
\address{School of Physics and Astronomy, Raymond and Beverly Sackler Faculty
of Exact Sciences,\\
Tel Aviv University, Tel Aviv 69978, Israel}
\date{\today}
\maketitle
\begin{abstract}

The effect of localized holes on the long-range antiferromagnetic
order in bilayer cuprates  is studied, by applying
the renormalization group to the appropriate classical non-linear sigma
model. The theory accounts quantitatively for the
magnetic phase diagram of
Ca doped YBa$_2$Cu$_3$O$_6$.

\end{abstract}

The parent compounds of high-$T_c$ superconductors, La$_2$CuO$_4$
(LCO) and YBa$_2$Cu$_3$O$_6$ (YBCO), are antiferromagnetic (AF) insulators.
Hole doping suppresses  rapidly the AF long-range order in both
compounds. In the monolayer 
La$_2$CuO$_4$ doped with strontium or with excess oxygen,
 the antiferromagnetic long range order  disappears at a
hole concentration $x$
close to 2\% \cite{KBB}, while in the bilayer YBCO the
critical concentration, $x_c$,
is about two times larger, $x_c\approx3.5\%$ \cite{CAM,NBB}.

In this paper we present the results of a theory, which explains
quantitatively the strong reduction of the transition temperature by
doping, and
the distinction between the magnetic phase diagrams
of LCO and YBCO.

It has been argued in Ref.~\cite{aharony} that holes residing
on the oxygen bonds in the CuO$_2$ planes generate an effective Cu-Cu
ferromagnetic coupling, which causes canting of the spins in the AF background.
The canting angle decays with distance similar to the potential of a 
magnetic dipole.
 Following Refs. \cite{glazman,CKA}, 
we describe the $\nu$-layered system by the reduced Hamiltonian (i.e.,
the Hamiltonian divided by the temperature $T$)
$ H= H_{0}+ H_{int}$.
Here $H_{0}$ is the non-linear-$\sigma$-model
 Hamiltonian in the renormalized classical region
\cite{CHN,Dots,YTC} for the undoped system:
\begin{equation}
 H_{0}=  {\rho_s\over2T}\int d{\bf r}\{\sum_{p=1}^{\nu}[
(\nabla {\bf n}^{(p)})^{2}+\gamma_{p}({\bf n}^{(p)}-
{\bf n}^{(p-1)})^2]\},
\label{ho}
\end{equation}
where $\rho_s$ is the in-plane spin
stiffness, and $\gamma_{p}$ represents the interlayer coupling,
with $\gamma_{1}=0$.

The Hamiltonian $ H_{int}$ describes the interaction of the (randomly
distributed in both planes) dipole  type impurities with the fields
${\bf n}^{(p)}$ \cite{glazman,CKA}. Denoting by ${\bf a}({\bf r}^{(p)}_{\ell})$
 the unit
vector directed along the doped bond at ${\bf r}^{(p)}_{\ell}$, and by
$M{\bf m}({\bf r}^{(p)}_{\ell})$ the corresponding dipole moment
(${\bf m}({\bf r}^{(p)}_{\ell})$ is a unit vector), 
 we have
\begin{equation}
 H_{int}={\rho_s\over T}\int d{\bf r}\sum_{i,p,p'}\beta_{pp'}{\bf f}_{i}^{(p)}
({\bf r})\cdot\partial_{i}{\bf n}^{(p')}.\label{hint}
\end{equation}
Here $\beta_{pp'}=\delta_{pp'} +\Gamma_p\delta_{p,p'-1}(1-\delta_{pp'})$,
$\Gamma_p\sim \gamma_p\ll1$, and 
${\bf f}^{(p)}_{i}({\bf r})=M\sum_{\ell}\delta ({\bf r}-{\bf r}_{\ell}^{(p)})
a_{i}({\bf r}^{(p)}_{\ell}){\bf m}({\bf r}^{(p)}_{\ell}).$
At temperatures lower than the  optical spin-wave gap (approximately 70 meV
for YBCO \cite{RBF}) only configurations
with ${\bf n}^{(1)} \approx{\bf n}^{(2)}$
dominate
the action \cite{Dots}, and the Hamiltonian  transforms to
\begin{equation}
 H=
{1\over 2t_{\nu}}\int d{\bf r}[(\nabla{\bf n})^2 +
2\sum_{i}{\bf f}^{(\nu)}_{i}
({\bf r})\cdot\partial_{i}{\bf n}],\label{hint1}
\end{equation}
with $t_{\nu}=T/\nu\rho_s$, and ${\bf f}_i^{(\nu)}\approx
\sum_{p}{\bf f}_{i}^{(p)}/\nu$.

The randomly distributed effective dipoles  develop dipole-dipole
interactions mediated by the AF spin background. \cite{aharony,glazman,CKA}
Therefore, at sufficiently low temperatures
 they either freeze in a random spin glassy way
or at least develop exponentially long range spin--glassy
correlations. Assuming that the range of these correlations is much larger than
that of the AF correlations, which in hole doped AF is finite
at all $T$ \cite{CKA}, justifies treating these moments as quenched.
The variables ${\bf r}^{(p}_{\ell}$ and ${\bf a}({\bf r}^{(p}_{\ell})$
are also quenched.
Denoting quenched averages
by $[...]$, we write
$\bigl [f_{i\mu}^{(\nu)}({\bf r})f_{j\mu'}^{(\nu)}({\bf r}')\bigr ]
=\lambda_{\nu}\delta_{\mu\mu'}\delta_{ij}
\delta ({\bf r}-{\bf r}')$,
with \begin{equation}
\lambda_{\nu}=M^2x/6\nu\equiv A_{\nu}x.
\label{lambda}
\end{equation}
The $\nu$-layer problem is thus mapped onto that of the monolayer
one \cite{glazman,CKA}, albeit with rescaled parameters $t_{\nu}$ and
$\lambda_{\nu}$. Both these parameters 
are marginal in the renormalization
group (RG)  sense. This allows to employ the RG methods to find
the effect of the dipole moments on a
two-dimensional (2D) AF. Generalising  the results
derived in Ref.~\cite{CKA} for LCO, with $\nu=1$,
the 2D correlation length obtained in the one-loop approximation is:
\begin{equation}
\xi_{2D}=C(T,x)\exp({2\pi/3\lambda_{\nu}}),~~~t_{\nu}<\lambda_{\nu},
\label{xit0}
\end{equation}
and
\begin{equation}
\xi_{2D}=C(T,x)\exp\Bigl(\frac{2\pi}{t_{\nu}}\Bigl[1-
\frac{\lambda_{\nu}}{t_{\nu}}
+\frac{\lambda_{\nu}^2}{3t_{\nu}^2}\Bigr]\Bigr), \lambda_{\nu}<t_{\nu}.
\label{xi1}
\end{equation}
Here $C(T,x)$ is a smooth function of its arguments, known 
for the bilayer ($\nu=2$) YBCO only at $x=0$ \cite{YTC}.

The 3D transition temperature $T_{N}(x)$
 of a system consisting of weakly
coupled planes may be deduced from the relation
\begin{equation}
\alpha \xi_{2D}^{2}(T_{N},\lambda ) \sim 1,
\label{3d}
\end{equation}
with $\alpha $ generated by an
interplane (interbilayer) exchange,
or some in--plane spin anisotropy.

Combining Eqs. (\ref{xit0}), (\ref{xi1}), and (\ref{3d}),
we  obtain the critical line $T_N(x)$.
 At $T_N(x)<\rho_s A_{1}x$, the
critical line is
expected to be practically
vertical, with $x_c$
 given by
\begin{equation}
x_c^{-1}=3A_{\nu}L(0,x_c),
\label{xc}
\end{equation}
where $L(T,x)=(1/4\pi)\ln(a^2/\alpha C^2(T,x))$, $a$ is the lattice constant.
 At smaller defect concentrations, 
i.e. for $T_N(x)$ higher than $\rho_s A_{1}x$, the critical line is
\begin{equation}
{T_N(x)\over T_N(0)}={A_{\nu}L(T_N,x)x\over 1-[1-3A_{\nu}xL(T_N,x)]^{1/3}}.
\label{Tn}
\end{equation}
\begin{figure}[h]
\epsfig{file=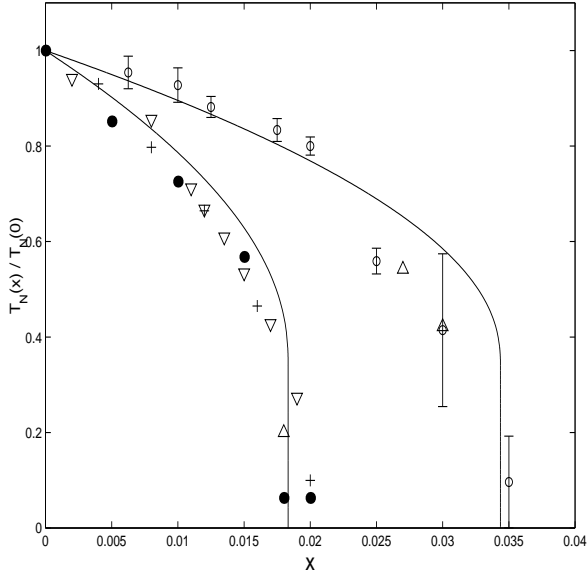,height=8cm,width=8cm}
\caption{
 $T_N(x)/T_N(0)$ versus hole concentration $x$. 
Full lines are theory (Eqs. (8),
(9)) for LCO (left line) and YBCO (right line). Symbols
are from experiments; LCO:
 $\bullet$ Ref.~[12], + Ref.~[13] 
$\bigtriangledown$ Ref.~[14], $\bigtriangleup$
 Ref.~[3], YBCO:
 $\circ$ Ref.~[2], $\bigtriangleup$ Ref.~[3].}
\end{figure}

The parameter $A_1\approx 20$ for LCO was estimated in Ref. \cite{CKA}
from the temperature dependence of $\xi$ in slightly doped samples.
It follows then from Eq. (\ref{lambda}) that $A_2$ for YBCO is
$A_2\approx 10$. Considering that the logarithmic factor $L$ does not
differ much for different cuprates,
we conclude that
  $x_c$ in YBCO
should be about two times larger than in LCO.
Since both $T_N$ and $x_c$ depend on $C(T,x)$ only
logarithmically, we neglect in what follows the temperature and concentration
dependence of $C(T,x)$. The prefactor $C(0,x_c)$ for LCO is approximately
 1.26 \AA \cite{CKA}, while $\alpha\approx 10^{-4}$ \cite{KBB}.
This gives $L\approx 0.91$.
 For YBCO we estimate $L$ from Eq. (\ref{3d})
for the undoped sample. With $T_N\approx 410$ K~\cite{SST},
 and $\rho_s\approx 200$ K (this follows from the data given in
Ref. \cite{SST}
and the quantum renormalization factors given in Ref. \cite{YTC}),
one gets $L\approx 0.97$.

Fig. 1 compares the theoretical critical lines, calculated from
Eqs. (\ref{Tn}) and (\ref{xc}) {\it without fitting parameters},
with the experimental data. The agreement is
very good for both LCO and YBCO, taking into account
 the uncertainties in the parameters $A$ and $L$,
obtained as explained above.

In summary, we applied the renormalization group to consider the
effect of localized holes (dipole-type impurities) on the magnetic
properties of Ca doped YBCO. We calculated the 2D correlation length
and obtained the phase diagram, in good agreement
with the experimental findings.

This project has been supported by the US--Israel Binational Science
Foundation.

\end{document}